# Superconductivity Phase Diagram of $Na_xCoO_2 \cdot 1.3H_2O$

R.E. Schaak[1], T. Klimczuk[1,2], M.L. Foo[1], and R.J. Cava[1,3]

[1]Department of Chemistry, Princeton University, Princeton NJ 08544
[2]Faculty of Applied Physics and Mathematics, Gdansk University of Technology, Narutowicza 11/12, 80-952 Gdansk, Poland
[3]Princeton Materials Institute, Princeton University, Princeton NJ 08540.

**Although the microscopic origin of the superconductivity in high $T_c$ copper oxides remains the subject of active inquiry, several of their electronic characteristics are well established as universal to all the known materials, forming the experimental foundation that all theories must address. The most fundamental of those characteristics is the dependence of the superconducting transition temperature on the degree of electronic band filling. Since the discovery of cuprate superconductivity in 1986 (1), the search for other families of superconductors that might help shed light on the superconducting mechanism of the cuprates has been of great interest. The recent report of superconductivity near 4K in the triangular lattice, layered sodium cobalt oxyhydrate, $Na_{0.35}CoO_2 \cdot 1.3H_2O$, suggests that superconductors related to the cuprates may be found (2). Here we show that the superconducting transition temperature of this compound displays the same kind of chemical-doping controlled behavior that is observed in the cuprates. Specifically, the optimal superconducting $T_c$ occurs in a narrow range of sodium concentrations, and therefore electron concentration, and decreases for both underdoped and overdoped materials, in analogy to the phase diagram of the cuprate superconductors. Our results suggest that detailed characterization of this new superconductor may help establish which of the many special characteristics of the cuprates is fundamental to their high $T_c$ superconductivity.**

Like the high $T_c$ superconductors, the $Na_xCoO_2 \cdot 1.3H_2O$ crystal structure (2) consists of electronically active planes (in this case, edge sharing $CoO_6$ octahedra) separated by layers (in this case, $Na_x \cdot 1.3H_2O$) that act as spacers, to yield electronic two-dimensionality, and also act as charge reservoirs. We have found that varying the Na content in $Na_xCoO_2 \cdot 1.3H_2O$ results in the same type of out-of-plane chemical doping control of in-plane electronic charge that is found for the cuprate superconductors. This is achieved by changing the Br concentration used in the deintercalation of the host material. (See caption of Fig. 1 for the synthesis procedure). Powder X-ray diffraction (XRD) patterns for the synthesized samples are shown in Fig. 1. The bromine-treated samples made with substoichiometric (0.5X) and stoichiometric (1X) bromine solutions consist primarily of a partially deintercalated, anhydrous, non-superconducting $Na_xCoO_2$ phase ($c \approx 11.2$ Å). A small amount of the hydrated superconducting phase $Na_xCoO_2 \cdot 1.3H_2O$ ($c \approx 19.6$ Å) is detectable by XRD for the 1X sample. Single phase, superconducting, fully hydrated $Na_xCoO_2 \cdot 1.3H_2O$ occurs for higher Br concentrations, with a small amount of $Na_xCoO_2$ in the 10X sample. Chemical analysis indicates that the sodium content of the resulting materials generally varies systematically in the samples prepared in different Br concentrations, over a range of $x = 0.26$ to 0.45 (Table 1). Thermogravimetric analysis of all the samples on very slow heating in oxygen showed that their behavior was identical to that reported previously for $Na_{0.3}CoO_2 \cdot yH_2O$ (3). The interlayer water content remains essentially const-ant, at approximately 1.3 per formula unit, despite differences in sodium content, as illustrated in the inset to figure 2. Figure 1 shows a noticeable shift in the positions of the 00*l* reflections for the fully hydrated $Na_xCoO_2 \cdot 1.3H_2O$ phases, yielding a systematic variation in the *c* axes of the unit cells (Table 1) from 19.43 Å for the $x = 0.45$ sample to 19.77 Å for the $x = 0.26$ sample. This increase in layer separation with decreasing sodium content for the hydrated superconducting phase is similar to that observed in the dehydrated $Na_xCoO_2$ phase (3,4). The *a* axis, reflecting the in-plane $CoO_6$ dimensions, is independent of Na content within the precision of our measurements.

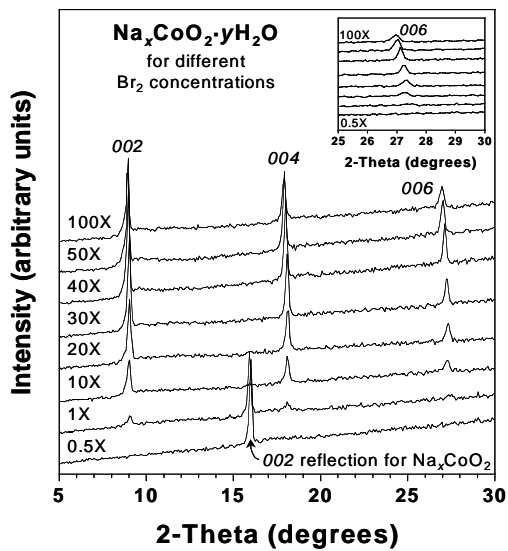

**Figure 1**. Powder X-ray diffraction patterns (Cu Kα radiation) for $Na_xCoO_2 \cdot yH_2O$ samples prepared using different concentrations of the bromine deintercalant. The inset shows an enlargement of the *006* reflections for each sample, highlighting the shift in the layer spacing as a function of sodium content. The $Na_xCoO_2 \cdot yH_2O$ samples were prepared by chemically deintercalating sodium from $Na_{0.7}CoO_2$ using bromine as an oxidizing agent (2,3). One-half gram of $Na_{0.7}CoO_2$ was stirred in 20 mL of a $Br_2$ solution in acetonitrile at room temperature for five days. Bromine concentrations representing substoichiometric (0.5X), stoichiometric (1X), and molar excess (10X – 100X) relative to sodium content were employed. ("1X" indicates that the amount of $Br_2$ used is exactly the amount that would theoretically be needed to remove all of the sodium from $Na_{0.7}CoO_2$.) The product was washed several times with acetonitrile and then water, and then dried briefly under ambient conditions. The sodium content of the phases was determined by the inductively coupled plasma atomic emission spectroscopy (ICP-AES) method. Very high Na diffusion coefficients facilitate homogenization of the Na contents of the samples at ambient temperature.

**Table 1**. Characterization of $Na_xCoO_2 \cdot yH_2O$ prepared by bromine deintercalation and hydration of $Na_{0.7}CoO_2$.

| Sodium content[a] ($x$ in $Na_xCoO_2 \cdot yH_2O$) | Bromine concentration | $a$ axis of $Na_xCoO_2 \cdot yH_2O$[b] (Å) | $c$ axis of $Na_xCoO_2 \cdot yH_2O$[b] (Å) | $T_c$[c] (K) |
|---|---|---|---|---|
| 0.45 | 0.5X | N/A | N/A | 2.0 |
| 0.40 | 1X | 2.823(3) | 19.43(2) | 2.0 |
| 0.32 | 10X | 2.825(2) | 19.52(2) | 2.1 |
| 0.33 | 20X | 2.823(2) | 19.58(1) | 2.2 |
| 0.32 | 30X | 2.822(2) | 19.58(1) | 3.0 |
| 0.30 | 40X | 2.823(2) | 19.69(2) | 4.3 |
| 0.29 | 100X | 2.819(3) | 19.77(2) | 4.0 |
| 0.26 | 50X | 2.821(2) | 19.77(2) | 2.4 |

[a]Sodium content determined by ICP-AES. The estimated error of analysis is ±0.02 per formula unit. [b]Determined by least squares refinement of powder XRD data, from 6-10 reflections between 5 and 60 degrees 2θ. [c]$T_c$s determined from the AC susceptibility data, from the extrapolation of the steepest slope of the M vs. T curves in Fig. 3 to M = 0.

Zero-field cooled DC magnetization data measured in a field of 5 Oe for selected samples are shown in the main panel of Fig. 2. The magnetizations at 1.8 K represent approximately 100% of the theoretical value expected for perfect diamagnetism. Such strong diamagnetic signals provide evidence for bulk superconductivity. Na concentration inhomogeneities in the samples are likely the primary source of rounding of the superconducting transitions. An important point revealed by the data in figure 2 is that $T_c$ for each sample is clearly different, indicating that differences in sodium content significantly affect the superconductivity of $Na_xCoO_2 \cdot 1.3H_2O$.

In order to characterize fully the dependence of $T_c$ on sodium content, we used AC susceptibility, which is more sensitive to weakly superconducting samples. Figure 3 shows the AC susceptibility for all the $Na_xCoO_2 \cdot yH_2O$ samples. For the multiple phase $x = 0.45$ and $x = 0.40$ samples, the $T_c$s are approximately 2.0 K (see table 1), and the diamagnetic AC signals are very small, consistent with their phase analysis by x-ray diffraction, which shows primarily non-superconducting, anhydrous $Na_xCoO_2$. In the $x = 0.40$ sample, the fully hydrated $Na_xCoO_2 \cdot 1.3H_2O$ phase accounts for approximately 15% of the sample, allow-ing us to estimate the maximum sodium content of the $Na_xCoO_2 \cdot 1.3H_2O$ phase to be approximately $x$ = 0.35. The data suggest that at its highest possible Na content, the $Na_xCoO_2 \cdot 1.3H_2O$ phase has a $T_c$ near 2K.

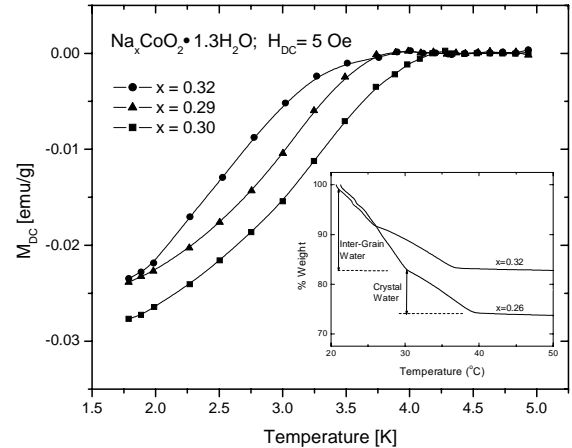

**Figure 2**. Zero field cooled DC magnetization (Quantum Design PPMS magnetometer, $H_{DC}$ = 5 Oe) for superconducting samples of $Na_xCoO_2 \cdot 1.3H_2O$ ($x$ = 0.29, 0.30, 0.32). The inset shows the loss in weight of single phase $Na_xCoO_2 \cdot 1.3H_2O$ ($x$ = 0.26, and 0.32) samples heated extremely slowly in $O_2$ (0.25 degrees per minute) illustrating the method by which we distinguish the amount of crystal water (the higher temperature weight loss) from the inter-grain water (the lower temperature weight loss). The change in weight that occurs on loss of crystal water is seen to be essentially the same in both low Na and high Na content materials.

All other samples are single phase sodium cobalt oxyhydrate with the crystal structure of the superconductor. The $x$ = 0.32 and $x$ = 0.33 samples yield slightly higher $T_c$s (between 2.1 and 2.2 K) and signals that are one to two orders of magnitude higher than the multiple phase $x$ = 0.40 sample. Single phase samples with sodium contents of $x$= 0.32, 0.30, and 0.29 display superconducting transition temperatures of 3.0 K, 4.3 K, and 4.0 K, respectively (Fig. 3). Significantly, the single phase sample with $x$ = 0.26 has a $T_c$ of only 2.4 K.

Fig. 4 shows the superconducting phase diagram of $Na_xCoO_2 \cdot yH_2O$. The variation of $T_c$ as a function of $x$ is shown in the main panel, and the magnitude of the DC magnetizations measured in an applied field of 5 Oe at 1.8 K, plotted on a logarithmic scale, are shown in the inset. These results clearly show that there is an optimal sodium composition for the occurrence of superconductivity ($x$ = 0.30) and that both the $T_c$ and quality of the superconductivity (inset, fig. 4) decrease at both lower and higher Na contents. As the sodium content increases between $x$ = 0.26 and $x$ = 0.35, the formal oxidation state of the Co decreases from 3.74+ to 3.65+. Consequently, $T_c$ varies dramatically with the degree of electronic doping of the $CoO_2$ planes, in analogy with the behavior observed in the cuprate superconductors. We note that the synthesis method we have employed, and other ambient temperature synthesis methods that might be used in this system, are likely to result in a distribution of sodium contents for each sample. If ideally uniform sodium content samples can be prepared, we expect that the superconducting "dome" shown in figure 4 may become more narrow in sodium content.

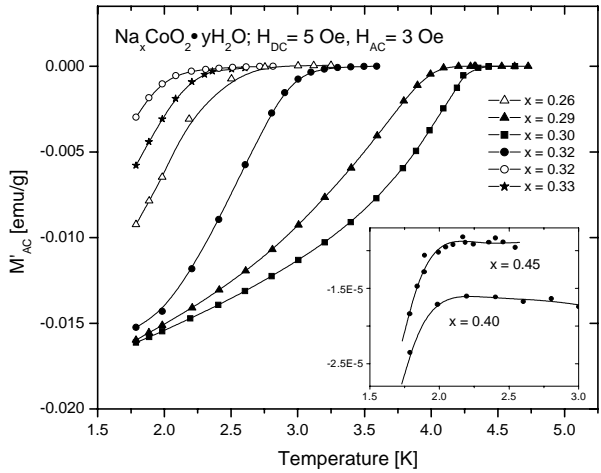
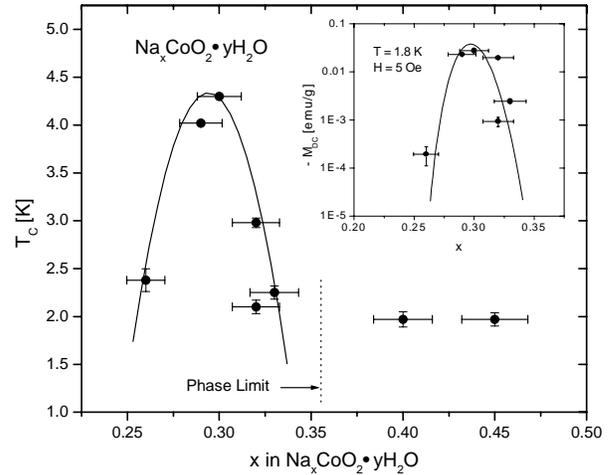

**Figure 3**. Zero field cooled AC magnetization for all superconducting $Na_xCoO_2 \cdot yH_2O$ samples. ($H_{DC}$ = 3 Oe, $H_{AC}$ = 5 Oe, $f$ = 10 kHz) Magnetization data for the weakly superconducting samples $x = 0.45$ and $x = 0.40$ are shown in the inset.

**Figure 4**. The superconducting phase diagram for $Na_xCoO_2 \cdot yH_2O$. Main panel: $T_c$ as a function of $x$ as determined from the AC susceptibility measurements in Fig. 3. Inset: the magnitude of the diamagnetic DC magnetization measured at 1.8 K in an applied DC field of 5 Oe, plotted on a logarithmic scale.

Preliminary correlation of the chemical doping due to the Na content and the true electronic doping state of the $CoO_2$ planes can be accomplished by electron counting in the context of electronic pictures already being developed for both the dehydrated $Na_xCoO_2$ and $Na_xCoO_2 \cdot 1.3H_2O$ phases (see, for example refs. 5-10). For $x = 0$, the formal Co oxidation state is $Co^{4+}$, with a $t_{2g}^5$ electron configuration in the low spin state. For $x = 1$, Co is formally 3+, with an electron configuration of $t_{2g}^6$ in the low spin state. Electronic structure calculations for $Na_{0.5}CoO_2$ (7) indicate that the $t_{2g}$ band would be completely filled at $x = 1$, and that an "ordinary" semiconductor is expected. However, it has been pointed out that the trigonal distortion of the $CoO_6$ octahedra in these structures may lead to splitting of the $t_{2g}$ band (6,11). The proposed $t_{2g}$ band splitting instead would result in a fully filled four-electron band, and a half filled two-electron band for $x = 0$. In this scenario, $Na_xCoO_2 \cdot 1.3H_2O$ for $x = 0$ bears a striking similarity to $La_2CuO_4$, the ground state for the cuprate superconductors, where electrons at the Fermi Energy also reside in a half filled two-electron band. It has not yet been determined whether this half filled state in the layered triangular lattice cobaltates at $x = 0$ gives rise to a Mott-Hubbard insulator as it does in the cuprates. In this scenario, then, each added Na above $x = 0$ in $Na_xCoO_2 \cdot 1.3H_2O$ results in the addition of one electron per cobalt to the half-filled band, and the opti-mal chemical doping level for superconductivity, $x = 0.3$, represents the addition of 0.3 electrons to the half-filled band per formula unit. Values of $x$ less than 0.3 would then represent underdoped materials, and values of $x$ greater than 0.3 would represent overdoped materials. More detailed characterization of the electronic state of the triangular cobaltates will be needed to determine just how closely the electronic analogies to the cuprates hold.

This work reveals several experimental findings that are critical for understanding the superconductivity of $Na_xCoO_2 \cdot 1.3H_2O$. The discovery of correlated maxima in both $T_c$ and diamagnetic shielding as a function of the Na content clearly establish the optimal chemical doping level for superconductivity. A fundamental similarity between the layered cuprate and layered cobaltate superconductors is seen in the decrease in $T_c$ for both the underdoped and overdoped materials. The optimal doping level for superconductivity appears to be relatively higher in the cobaltates than in the cuprates. Observations of unusual electronic properties in the host material $Na_{0.7}CoO_2$ itself (11,12) suggest that coupled spin and charge dynamics may be implicated in the superconductivity. Observations that the lower hydrates with closer $CoO_2$-$CoO_2$ interplanar distances are not superconducting above 2K (3), and that $T_c$ decreases under pressure (13) indicate that the two dimensional character of the structure is important. Though the intrinsically complex materials chemistry of the $Na_xCoO_2 \cdot 1.3H_2O$ superconductor makes it difficult to characterize, we believe that potentially fruitful comparisons to the cuprates, and the fact that this compound may represent the literal embodiment of Anderson's original proposal for the RVB (resonating valence bond) state (14), make it highly worthy of further study.

## Acknowledgements

This work was supported by the US National Science Foundation, Division of Materials Research, and the US Department of Energy, Division of Basic Energy Sciences. T. Klimczuk would like to thank The Foundation for Polish Science for support.